\documentclass[12pt,showpacs,amssymb,aps,pra,tighten,amsmath]{revtex4-1}
\usepackage{graphicx}

\begin{document}

\title{Maxwell-Drude-Bloch dissipative few-cycle optical solitons}

\author{Nikolay N. Rosanov,$^1$ Victor V. Kozlov,$^{2,3}$ and Stefan Wabnitz$^2$}
\affiliation{$^1$Institute of Laser Physics, Vavilov State Optical Institute,
St.-Petersburg, 199034, Russia.\\
$^2$Department of Information Engineering, Universit\'{a} di
Brescia, Via Branze 38, 25123 Brescia, Italy.\\
$^3$Fock Institute of Physics, St.-Petersburg State University,
Petrodvoretz, St.-Petersburg, 198504, Russia}

\begin{abstract}
We study the propagation of
few-cycle pulses in two-component medium consisting of nonlinear
amplifying and absorbing two-level centers
embedded into a linear and conductive host material. First we present a 
linear theory of propagation of short pulses in a purely conductive material, 
and demonstrate the diffusive behavior for the evolution of the low-frequency 
components of the magnetic field in the case of relatively strong conductivity. 
Then, numerical simulations carried out
in the frame of the full nonlinear theory involving the Maxwell-Drude-Bloch 
model reveal the stable
creation and propagation of few-cycle dissipative solitons under 
excitation by incident femtosecond optical pulses of relatively high energies. 
The broadband losses that are introduced by the medium conductivity represent 
the main stabilization mechanism for the dissipative few-cycle solitons.
\end{abstract}

\pacs{\hbox{}\\
42.65.Tg Optical solitons; nonlinear guided waves  \\
42.65.Re Ultrafast processes; optical pulse generation and pulse compression \\
42.50.Md Optical transient phenomena: quantum beats, photon echo, free-induction
decay, dephasings and revivals, optical
nutation, and self-induced transparency  \\
42.50.Nn Quantum optical phenomena in absorbing, amplifying, dispersive and
conducting media; cooperative phenomena in
quantum optical systems }
\maketitle

\section{Introduction}
Few-cycle optical pulses with durations of only a few periods of the optical
radiation \cite{BK,DGC} dramatically expand the
horizon of modern optics, and offer new applications in metrology, ultrafast
spectroscopy, and material processing. Such pulses are of
particular relevance to the field of extreme nonlinear optics, where
electromagnetic fields of enormously large intensities are
required \cite{MTB}. Optical soliton effects may play an important role in the
process of generation of few-cycle pulses,
as well as in the course
of their propagation through nonlinear media. As a matter of fact, the
appropriate description of
few-cycle pulse generation and propagation reveals new, not previously
anticipated
physics, in parallel calling for the abandonment of most of the approximations
which are of standard use in nonlinear optics,
and for the development of new approaches to the treatment of the interaction of
spectrally broadband radiation with matter.

To the best of our knowledge, the first example of conservative few-cycle
soliton was found in the context of nonreduced
Maxwell-Bloch equations by Bullough and Ahmad in Ref.~\cite{BA_71}. That soliton
was of the video type, that is with zero carrier frequency, with an hyperbolic
secant shape for the total electric field of the pulse and not its envelope.
More recent studies,
see e.g. \cite{KSAHL,AVB}, have considered a new class of conservative few-cycle
solitons which are the solutions of a properly
generalized nonlinear Schroedinger equation, without the assumption of a slowly
varying field envelope as well as the
approximation of unidirectional propagation. The dispersive properties of the
medium that was modeled in those studies imply the
singularity of the dielectric constant at zero frequency, which leads to the
area constraint $S_E \equiv \int_{-\infty}^{+\infty} dt\, E=0$
that was imposed on the electric field $E$ of any localized field distribution 
such
as the soliton.
This area constraint may appear to be violated in real-world experiments, 
whenever
the
zero-frequency component of the field is generated in a coherent source either
spontaneously or deterministically, e.g., via optical rectification in a medium
with quadratic nonlinearity. It is then natural to undertake the study of the
dynamics of low-frequency Fourier
components of the electromagnetic field in situations where the electric field 
area constraint
is lifted: this is one of the goals that we will pursue in our present study.

Additionally, the above mentioned studies were devoted to the properties of
conservative solitons, hence the effects of medium gain and loss were not
considered. However, losses (as well as the gain which is necessary to balance
them) come to the forefront whenever one sets the goal of generating few-cycle
pulses, and possibly propagating them
over long distances. Moreover, as well known, in active media it is exactly the
frequency-dependence of gain (and loss) which determines the utmost degree of
achievable pulse compression. Soliton type effects in combination with losses
and gain give rise to localized field structures known as dissipative solitons,
which are of primary interest in our study. Indeed, in our work we will exploit
a full-scale analysis of the interaction of a strong few-cycle pulse with a
dissipative medium consisting of a two-component (i.e., amplifying and
absorbing) ensemble of two-level doping centers (atoms or quantum dots),
embedded in a conductive host. In doing that, we will reject the usual
approximation of
slowly varying envelopes, as well as the unidirectional approximation and the
approximation of zero "electric area" $S_E=0$. In short, in this paper we are
going to investigate the formation of few-cycle dissipative solitons within the 
framework of
the non-reduced Maxwell-Drude-Bloch equations.

Few-cycle dissipative solitons based on the phenomenon of the McCall and Hahn
self-induced transparency
\cite{McCall_67,McCall_69} were first theoretically studied in
Refs.~\cite{VRSFW,VRS_06}.
In those studies, the propagation medium was composed of two
sorts of resonant atomic systems -- absorbing and amplifying. The idea behind
the formation of dissipative solitons is the
following. In the absence of dissipative factors, such as loss and gain, a
two-level resonant atomic system supports the formation of
a conservative soliton. As well known, such solitons form a family
of solutions, where the soliton peak amplitude is a
continuously varying parameter. The temporal
duration of these solitons is inversely proportional to their amplitude. If one
now allows for the existence of a weak linear gain (a
small dissipative factor), the propagation dynamics leads to a continuous
transition with distance to solitons within the family with progressively higher
amplitudes, and
correspondingly shorter durations.
In fact, the soliton amplitude grows exponentially with the propagation length,
and its time duration also shortens exponentially: both factors work together
and result in the temporal collapse of the soliton. At first sight, such
shortening would be welcome as it produces high power, few-cycle pulses.
However, such behavior is not realistic in practice, as it does not account for
nonlinear, dispersive and absorptive mechanisms that work against the collapse,
and that are inherently present in any real medium. Moreover, when only linear
gain is present, (bright) localized structures appear to be unstable against
amplification of perturbations on the tails of their field.
When taking into account the nonlinearity of a two-level gain
medium, which is supposed to be coupled with a relaxing two-level absorber, one
observes the appearance of a set of localized few-cycle pulses with stationary
shapes and exhibiting a discrete set of velocities, see Ref.~\cite{VRSFW}.
Later on in Ref.~\cite{VRS_06} it was shown that these
localized pulses are not stable, since their collapse is not counterbalanced by
the nonlinearity of the gain. Nevertheless,
it was also predicted that stabilization of the soliton is possible by
introducing a third level into the model of the absorbing medium.
In this way, stable half-cycle (or video-) dissipative solitons were obtained
and studied in detail, see Refs.~\cite{VRS_06, RSV_07,RSV_08}. More recently,
few-cycle dissipative solitons (containing several oscillations of the field)
were found in Ref.~\cite{VRS_09}
for the model of amplifying and
absorbing homogeneously as well as inhomogeneously broadened quantum dots,
embedded into a quartz host material. In that work, the
host matrix was modeled by means of a three-level system with two separate
resonance frequencies located in the infra-red and in the
ultra-violet spectral domains, respectively. In this way, the linear (absorptive 
and dispersive) properties of
the host medium in its transparency region could be reproduced with high
accuracy.
For this model, it was of paramount importance (from the standpoint of practical
applications) to observe that the few-cycle dissipative solitons could be
excited by means of a much longer femtosecond pulse incident on the medium. Such
soliton generation mechanism exhibits a threshold behavior: namely,
low-intensity pulses
disperse and decay, whereas the evolution of above-threshold pulses results in
the formation of dissipative solitons. Note that the
(nonstationary) propagation of a few-cycle pulse in a two-level amplifier was
studied in Ref.~\cite{T_OC}, however without
counterbalancing the gain by the nonlinearity as well as the linear or nonlinear
losses of the absorber.
Therefore the type of dissipative solitons which are of present interest to us
cannot be formed in the configuration that was considered in
Ref.~\cite{T_OC}.

All of the above-mentioned studies
assume that the response of the medium to the
electromagnetic field is solely described in terms of the medium polarization. 
However,
according to the Lorentz electrodynamics of continuous media, the medium 
interaction
with radiation is governed not only by the response of bound charges
(electrons), i.e. medium polarization, but also by the response of free
electrons, i.e., the electric current. In this paper we present a significant
step forward in our studies of few-cycle dissipative optical solitons, by
proposing a new approach that is based on the full Lorentz model,
which eventually leads us to the formulation of the Maxwell-Drude-Bloch model.
We demonstrate that such a model is appropriate for treating the medium of our
interest, namely an ensemble of active (amplifying) and passive (absorbing)
quantum dots embedded into a semiconductor, off-resonance host matrix. The
semiconductor host material is characterized by an
appreciable value of the electrical conductivity, which may be most easily
regulated by a proper choice of the concentration of dopants. We
demonstrate that conductivity provides a stabilization mechanism for dissipative
solitons, even in the case where simple two-level models
are used to describe active and passive quantum dots. As a matter of fact, the
role of conductivity is two-fold.
First of all, conductivity provides broadband
(although tilted with frequency) linear losses, which are of vital importance
for the stabilization of solitons. The second property (that is not related with
soliton stabilization) of conductivity is that it leads to high reflectivity for 

low-frequency field components in the
experimental situation which is of interest for us,
namely an electromagnetic pulse that is
incident at the boundary with a conductive medium. As a matter of fact, in
this situation
the Fresnel reflectivity is unity for the zero-frequency
component of the field. The role of conductivity has been so far
under-appreciated
in both linear and nonlinear optics, albeit for a good reason,
namely because the carrier frequency of a laser pulse is so high and its
spectrum is so narrow (with respect to the carrier frequency)
that the zero-frequency
component of the optical field has practically zero intensity.
However, the situation changes when the pulse becomes as short as a few optical
oscillations,
or when it represents a video-pulse. In such situations,
the zero-frequency component cannot be ignored anymore: the conductivity of
semiconductor materials strongly affects the spectrum of both the propagating as
well as the reflected pulse. As a consequence, we need to generalize the usual
Maxwell-Bloch approach which describes the interaction of the electromagnetic
field with bound electrons, in order to include its
interaction with the medium free electrons. The dynamics of the latter is
modeled by means of the Drude equation, so that the whole light-matter
interaction picture must be based on the simultaneous solution of the
Maxwell-Drude-Bloch equations (some relevant studies concerning the
photovoltaic and photorefractive effects are presented in Ref.~\cite{SF}).

Since the few-cycle dissipative solitons that we are going to investigate in
this work
have their roots back in the phenomenon of self-induced transparency, a related
and significant subject of our
discussion is the notion of the pulse area. For pulses of slowly varying
envelopes, the pulse area was introduced as the time
integral (between minus and plus infinity) over the pulse envelope. Such pulse
envelope area
obeys the celebrated McCall and Hahn Area Theorem, see \cite{McCall_69}.
However, the notion of pulse envelope is no longer appropriate for few-cycle
pulses: as a consequence, the Area Theorem is no longer valid,
see Refs.~\cite{H_98,T_OE}. For few-cycle pulses, instead of the envelope area,
it is appropriate to consider
the electric field area, which is defined as the time integral
with infinite limits of the electric field itself (and not its envelope). A
simple equation governing evolution of the electric field area
in a rather general medium was derived in Ref.~\cite{R_09}. Physically, the
electric field area represents the
zero-frequency component of the Fourier spectrum of the electromagnetic field.
Such a component is of no practical
importance for narrowband optical pulses, since in this case we may suppose a
vanishing zero-frequency field amplitude. Nevertheless, the electric field area 
may
gain a valuable meaning in the case of ultra-broadband (few-cycle) pulses, where
the
zero-frequency component of the total field may no longer be negligible. In this
work, we also
introduce the notion of the area of the magnetic field, which is
defined as the integral of the magnetic field over the spatial coordinate. In
practice, such definition is more appropriate than the definition of the
electric field area, since the evolution of the magnetic field area may be
easily numerically computed when solving realistic initial-value problems, i.e.,
whenever the electromagnetic field is specified along the spatial coordinate,
and time is the evolution coordinate.

In summary, the primary goal of this paper is the study of the propagation of
few-cycle pulses
through a two-component medium consisting of (nonlinear)
amplifying and absorbing two-level centers (atoms or quantum dots),
embedded into a (linear) host material that exhibits an appreciable value of
electrical
conductivity. Physically, we think of the host material as a semiconductor,
since it can be manufactured with a high concentration of
electrons in the conduction band. As appropriate candidates for the two-level
absorbing and amplifying centers, we envisage the use of two different types of
quantum dots, which possess huge dipole moments, so that one may avoid material
breakdown by the otherwise highly intense field of the few-cycle pulse (note
that the Rabi frequency can be as high as the optical frequency). The large
nonlinearity of the quantum dots permits us to neglect the nonlinearity of the
host semiconductor material. For the sake of simplicity, we also neglect in this
work the dispersive properties of the host material, which can be introduced
following the approach of Ref.~\cite{VRS_09}, and whose presence does not affect
the generality of the present results. We start our paper with a linear
theory of propagation of short pulses in a purely conductive material (i.e.,
with no embedded two-level systems), a topic which has not yet
received much attention in optics. Next we proceed to develop the nonlinear 
theory
of propagation of few-cycle pulses in a conductive material that is doped with
resonant amplifying and absorbing two-level doping centers. Finally, we conclude
by providing extensive numerical simulation results demonstrating the stable
formation and propagation of few-cycle dissipative solitons
in the framework of the fully nonlinear Maxwell-Drude-Bloch model.

\section{Maxwell equations and the electromagnetic area}
Let us consider a one-dimensional electromagnetic pattern of finite temporal and
spatial extension. This can be for example a pulse
propagating in a single-mode optical waveguide, in situations where the
associated waveguide dispersion may be neglected,
namely whenever the central part of the pulse spectrum is far from the mode
cut-off frequency. In this case, it is possible to derive separate conservation
laws for both the electric as well as the magnetic field area. Two of the
Maxwell
equations are relevant to us here. They read as
\begin{eqnarray}
&& \nabla\times{\textbf E}=-\frac{1}{c}\frac{\partial {\textbf B}}{\partial t}\,
,
\label{1}\\
&& \nabla\cdot {\textbf B}=0\, .
\label{2}
\end{eqnarray}
Here ${\textbf E}$ and ${\textbf B}$ are the strength of the electric field and
of the magnetic induction, respectively, and $c$ is the speed
of light in vacuum.  In a one-dimensional geometry, the field strength only
depends on one (longitudinal) coordinate, say, $z$ and
on time $t$. Therefore, as it follows from Eq.~(\ref{2}), the $z$-component
$B_z$ of the magnetic field is zero. Then, Eq.~(\ref{1})
takes the form
\begin{equation}
-\frac{\partial E_y}{\partial z}{\textbf e}_x+\frac{\partial E_x}{\partial
z}{\textbf e}_y=-\frac{1}{c}\frac{\partial{\textbf B}_\perp}{\partial t}\, ,
\label{3}
\end{equation}
where ${\textbf e}_{x,y}$ are unit vectors in the Cartesian frame orthogonal to
the propagation variable $z$, and
${\textbf B}_\perp =(B_x,\, B_y)^T$.

Next we define the magnetic area of the pulse as
\begin{equation}
{\textbf S}_B(t)=\int_{-\infty}^{+\infty}dz\, {\textbf B}(z,t)\, .
\label{4}
\end{equation}
Note that the longitudinal component of this vector is zero.
Then, integrating Eq.~(\ref{3}) over the longitudinal coordinate $z$,
one obtains the equation
\begin{equation}
\frac{d}{dt}{\textbf S}_B=0\, ,
\label{5}
\end{equation}
which is nothing but the conservation law for the magnetic area. That is, the
amplitude of the zero-frequency (or dc)
component of the Fourier
spectrum of the magnetic field strength does not change with time. This law
separately applies to both polarization
eigenmodes of the field. Similarly, by introducing the electric field area of
the electromagnetic pulse as
\begin{equation}
{\textbf S}_E(z)=\int_{-\infty}^{+\infty}dt\, {\textbf E}_\perp (z,t)
\label{6}
\end{equation}
we get
\begin{equation}
\frac{d}{dz}{\textbf S}_E=0\, ,
\label{7}
\end{equation}
after integration of Eq.~(\ref{3}) over time. The interpretation of the
conservation law for the electric area
as in Eq.~(\ref{7}) is, similar to the case of the magnetic area, the
time-invariance of the dc component of the electric field.
Note that in the course of the above derivations we did not explicitly use any
material equation. Nevertheless, we
supposed that the electric and magnetic fields vanish at the pulse tails.
Generally speaking, this only happens in the
presence of a relaxation in the material response to the electromagnetic field.
In simplified models, whenever such relaxation is neglected, the above discussed
area conservation properties do not hold. Unlike
the area theorem of self-induced transparency, which was derived for a
conservative two-level absorber, the conservation law for the electric and
magnetic areas, see Eqs.~(\ref{5}) and (\ref{7}), directly follows from
Maxwell's equations, without the necessity to provide any specification about
the medium (that can even be inhomogeneous) where wave propagation occurs.

Next, let us involve in our considerations another Maxwell equation which
provides a link to the optical properties of the material. We
suppose the medium to be electrically neutral, i.e., with electric charge
density $\rho =0$, and the electromagnetic field to
be transverse, i.e.
the longitudinal components of the electric field ${\textbf E}$, of the electric
displacement ${\textbf D}$, and of the current density
${\textbf j}$, all vanish. In the adopted one-dimensional geometry, we arrive to
the wave equation
\begin{equation}
\frac{\partial^2{\textbf E}_\perp}{\partial z^2}-\frac{4\pi}{c^2}\,\frac{\partial
{\textbf j}_\perp}{\partial t}
-\frac{1}{c^2}\,\frac{\partial^2{\textbf D}_\perp}{\partial t^2}=0\, .
\label{8}
\end{equation}

Since we are also interested in studying the dynamics of low-frequency radiation
in the conductive medium, it is instructive to analyze the
behavior of pulses whose spectra lie in the vicinity of zero frequency. Then we
may neglect the last term in
Eq.~(\ref{8}) and also use the Ohm law for the current,
\begin{equation}
{\textbf j}=\sigma_0{\textbf E}\, ,
\label{9}
\end{equation}
where $\sigma_0$ is the static conductivity of the medium. In the above limiting
situation, the wave equation is transformed to the one-dimensional parabolic
equation
\begin{equation}
\frac{4\pi\sigma_0}{c^2}\,\frac{\partial {\textbf E}_\perp}{\partial
t}=\frac{\partial^2{\textbf E}_\perp}{\partial z^2}\, .
\label{10}
\end{equation}
Similar equation holds for the magnetic induction ${\textbf B}_\perp$, see e.g.
\cite{Jackson}.
Such type of equation is quite unusual for describing optical phenomena.
Indeed, Eq.~(\ref{10}) is mostly applicable to video-pulses rather than to
narrowband or even
few-cycle optical pulses, also in the THz region.
On the other hand, Eq.~(\ref{10}) is widely known in the theory of heat
conductivity and diffusion, \cite{CJ} and also in the description of 
quasi-static
electromagnetic fields, Foucault currents and skin effect in conductors 
\cite{LL}.

The above derived parabolic equation is valid whenever the
following inequalities are satisfied:
\begin{equation}
\omega\ll\nu\, ,\quad \omega\ll 4\pi\sigma_0\, ,\quad \omega\ll
4\pi\sigma_0/\vert\epsilon_0\vert\, ,
\label{11}
\end{equation}
where $\nu$ is the effective frequency of collisions of free electrons with ions
and atoms, and $\epsilon_0$ is the
static component of the dielectric permittivity, possibly including a nonlinear 
contribution. Since for most situations $\vert\epsilon_0\vert \sim 1$,
the last inequality in Eq.~(\ref{11}) can be omitted. If we come back to the
original Maxwell's equations and estimate the value of the
electric field strength versus the magnetic field strength in such regime, we
arrive at an interesting inequality
\begin{equation}
E\sim\sqrt{\frac{\omega}{4\pi\sigma_0}}H\ll H\, .
\label{12}
\end{equation}
Namely, the electromagnetic field of a low-frequency pulse in a highly
conductive material mainly consists of the magnetic field,
in contrast to dielectrics where magnetic and electric fields are of comparable
strength.

In practical situations involving relatively large observation times, 
the detailed form of the initial and boundary conditions
for the diffusion equation (\ref{10})
may become irrelevant, and we may consider the medium as infinite. Let us
briefly overview the main characteristics of the field diffusion process in an
infinite medium. As a field, we take the magnetic component ${\textbf B}$ of the
electromagnetic field. This choice is dictated at first by the dominance of the 
magnetic field over the electric field as expressed by Eq.~(\ref{12}). Moreover, 
as we shall see in the following, this choice has interesting consequences in 
our discussion of the evolution of the magnetic area, and leads to a better 
(when compared with the electric field)
correlation with the analytically tractable case of ``transparent" boundary
conditions.  For an infinite medium, we may introduce $n$-th
momentum of the field ($n=0,\, 1,\, 2,\, \dots$) according to the formula
\begin{equation}
{\textbf S}_n(t)=\int_{-\infty}^\infty dz\, z^n {\textbf B}_\perp (z,\, t)\, .
\label{12_1}
\end{equation}
Note that zeroth momentum coincides with the definition of the magnetic area
given in Eq.~(\ref{4}). Substituting these definitions into the diffusion 
equation
(\ref{10}), we get for first three momenta
\begin{eqnarray}
&& \frac{d{\textbf S}_0}{dt}=\frac{d{\textbf S}_1}{dt}=0\, ,
\label{12_2}\\
&& \frac{d{\textbf S}_2}{dt}=2D{\textbf S}_0\, ,
\label{12_3}
\end{eqnarray}
where $D=c^2/4\pi\sigma_0$ is the diffusion coefficient. The conservation of the
zeroth momentum is a consequence of a more general area theorem (\ref{5})
for the magnetic field. The conservation of the first momentum means that the
centers of the magnetic distributions defined for corresponding polarizations as
$Z_x=S_{1x}/S_{0x}$ and $Z_y=S_{1y}/S_{0y}$ do not move in the course of the
diffusion. In its turn, the evolution of the second momentum reflects the 
broadening
of the magnetic field distribution with time. Namely, the width of the 
distribution grows larger in proportion to $(t-t_0)^{1/2}$.
Bearing in mind the conservation of the zeroth momentum, we may conclude that 
the
amplitude of the distribution correspondingly decreases as $(t-t_0)^{-1/2}$. 
When
compared with the standard exponential decay of the field in an absorbing
medium, in this case the dynamics of the magnetic field evolution is extremely 
slow (power-law decay).  The above discussed behavior of the first three momenta 
of the field is typical of any diffusion process in
an infinite medium. For asymptotically large evolution times $t$, the memory of 
any particular initial
condition at $t=0$ dies out, and the field evolves self-similarly as
\begin{equation}
{\textbf B}_\perp \propto \frac{1}{[D(t-t_0)]^{1/2}}\,
\exp\left[ -\frac{z^2}{4D(t-t_0)}\right]\, ,
\label{12_4}
\end{equation}
where $t_0$ is arbitrary an instant in time that is asymptotically far from 
$t=0$.

More relevant to our present study of the propagation of pulses incident on the 
boundary with a
conductive material is the consideration of the corresponding boundary-value 
problem. 
Let us consider first the analytically
tractable case of ``transparent" boundary, i.e. a  boundary which is transparent 
for
low-frequency radiation. We assume that the initial pulse that is incident on 
the boundary with
the semi-infinite medium at $t=0$ has already diffused sufficiently deep into 
the medium, so
that the value of ${\textbf B}$ at the boundary (located at $z=0$) has become 
vanishing small.
In this asymptotic limit, the field evolves self-similarly as
 \begin{equation}
{\textbf B}_\perp \propto \frac{1}{[D(t-t_0)]^{3/2}}\,
\exp\left[ -\frac{z^2}{4D(t-t_0)}\right]
\label{12_5}
\end{equation}
for $z\ge 0$. In order to accomodate the definitions of the momenta given in 
Eq.~(\ref{12_1})
to the case of a semi-infinite medium, we change in these expressions the lower 
integration limits to
$0$, and supply such re-defined momenta by the superscript $(tr)$ (standing for 
transmitted).
Then, as a result of the escape of the field into vacuum the zeroth order 
momentum (magnetic area
of the transmitted field) is no longer conserved. As a matter of fact, for 
asymptotically large times the transmitted magnetic area decays as ${\textbf 
S}_0^{(tr)}\propto (t-t_0)^{-1/2}$. 
However, the first order momentum is still conserved,
therefore the centers of gravities of the field distributions $Z_x$ and $Z_y$ 
penetrate deeper and
deeper into the medium as time evolves [$\propto (t-t_0)^{1/2}$]. However, the 
speed of this
motion slows down with time with a the rate proportional to $(t-t_0)^{-1/2}$. 
Although the real boundary
conditions in our present model have a more complicated form, nevertheless in 
the next section we will make use of the above discussed asymptotic diffusive 
time-dependency of the low-frequency components of the magnetic field, in order 
to
demonstrate its diffusive behavior in
a highly conductive medium.

Leaving the parabolic equation (\ref{10}) aside for the time being, let us
return to the wave equation (\ref{8}).  We do not assume
anymore that the conductivity acquires extreme values. Therefore both material
terms in the wave equation should be considered on
equal footing. For low-intensity pulses a linear propagation theory is readily
applicable. The evolution of each Fourier component of
the electromagnetic pulse can be represented by a running plane wave
\begin{equation}
{\textbf E}_\perp =\mbox{Re}\left\{ {\textbf E}_\omega \exp [i(kz-\omega
t)]\right\}\, .
\label{13}
\end{equation}
Similar decompositions take place for the current and the electric induction:
\begin{eqnarray}
&& {\textbf j}_\perp =\mbox{Re}\left\{ {\sigma (\omega )\textbf E}_\omega \exp
[i(kz-\omega t)]\right\}\, .
\label{14}\\
&& {\textbf D}_\perp =\mbox{Re}\left\{ {\varepsilon (\omega )\textbf E}_\omega
\exp [i(kz-\omega t)]\right\}\, .
\label{15}
\end{eqnarray}
where both the frequency-dependent complex-valued conductivity $\sigma (\omega
)$ and the dielectric permittivity
$\varepsilon (\omega )$ depend on the model in use.  Upon substitution of the
decompositions (\ref{13})-(\ref{15}) into the wave
equation (\ref{8}), one obtains the dispersion relation of the wave number in
the form
\begin{equation}
k(\omega )=\pm\sqrt{\frac{\omega^2}{c^2}\,\varepsilon (\omega
)+i\frac{4\pi\omega }{c^2}\,\sigma (\omega )}\, .
\label{16}
\end{equation}
The two signs before the square root correspond to the two opposite directions
of wave propagation. Since we are interested in soliton
dynamics, our model will be a nonlinear one. However, the linear theory is still
valuable because it allows us to
check the stability of a dissipative soliton against the undesirable
amplification of low-intensity field perturbations to the soliton tails. As a
matter of fact, the stability of any (bright) localized nonlinear field 
structure or
pulse requires the trivial
solution (${\textbf E}=0$) of the wave equation (\ref{8}) to be also stable.
Otherwise, the pulse tails will grow up exponentially in their propagation, so 
that the localized
structure is also unstable. Formally, the stability of the trivial solution 
means
that the small-signal gain coefficient does not exceed the small-signal
absorption
coefficient at any frequency $\omega$. As soon as we will complete the
description of our model and define the functions
$\sigma (\omega )$ and $\varepsilon (\omega )$, we will come back to this
problem and check the stability of the trivial solution ${\textbf E}=0$.

\section{The Drude equation and the Maxwell-Drude model}
Let us start the introduction of the material equations with an equation for the
electric current. According to the Lorentz macroscopic
electrodynamic theory, a medium contains two types of charges (electrons) --
bound and free, \cite{LL}. The dynamics of
bound charges is described by a (generally, nonlinear) polarization, whose
evolution is governed by the equations for the
density matrix (or Bloch equations). In our case, these equations describe two
types of two-level atom-like systems -- amplifying
and absorbing centers. The corresponding equations will be presented in the next
section. Here we concentrate on the dynamics of free charges, and follow the
equation of motion for the current as it was proposed by Drude. This equation
reads as
\begin{equation}
\frac{d{\textbf j_\perp}}{dt}+\nu{\textbf j}_\perp =\frac{\omega_p^2}{4\pi} \,
{\textbf E}_\perp\, .
\label{17}
\end{equation}
Here $\omega_p^2=4\pi e^2 N_e/m_e$ is the square of the plasma frequency, where
$e$, $m_e$, and $N_e$ are the electron
charge, the electron mass and the concentration of free electrons, respectively.
Let us remind
here that $\nu$ is the frequency of collisions of free
electrons with ions and atoms. This collisional frequency  plays the role of a
relaxation constant. Collisions damp the dynamics
which is dictated by the electric field.  The Drude model is valid not only for 
plasmas
and metals but also for semiconductors. For the latter,
 it is however
necessary to change the free electron mass $m_e$ to the effective mass
$m_{\mbox{eff}}$ of the carriers in the conduction band.
In a low-frequency limit ($\omega\ll\nu$), the collisional term dominates in the
left-hand side of Eq.~(\ref{17}), and we thereby
recover the static Ohm law, see Eq.~(\ref{9}), where the value of static
conductivity is given by expression
\begin{equation}
\sigma_0=\frac{\omega_p^2}{4\pi\nu}\, .
\label{18}
\end{equation}
Whenever necessary, the Drude equation (\ref{17}) may be generalized by
including additional terms that are nonlinear
in the electric field, as for instance it is
relevant when considering photovoltaic phenomena \cite{SF}.

The model of conductivity that is supplied by the Drude equation (\ref{17})
assumes the following form of the frequency-dependent conductivity
entering in Eq.~(\ref{14}):
\begin{equation}
\sigma (\omega )=\frac{\omega_p^2}{4\pi}\,\frac{1}{\nu -i\omega}\, .
\label{19}
\end{equation}
Since the resonance frequency of (not yet modeled) narrowband resonant two-level
centers is relatively far from the zero
frequency, at low frequencies the conductivity term dominates under the radical
in Eq.~(\ref{16}). In this limit, the only effect of the
medium on the low-frequency components is their (frequency-dependent) decay in 
the
course of the propagation.

From here on, we suppose for simplicity that the state of polarization of
radiation incident on the medium interface is linear:
${\textbf E}_\perp =E{\textbf e}_x$, and therefore ${\textbf B}_\perp =B{\textbf
e}_y$ and ${\textbf j}_\perp =j{\textbf e}_x$. In this case, the wave equation
(\ref{8}), the parabolic equation (\ref{10}), and the Drude equation (\ref{17})
are expressed in a scalar form.

It is instructive to consider the Maxwell-Drude model separately from the
equations for the bound charges (i.e. the Bloch equations),
because the influence of resonant two-level centers is negligible at low
frequencies. Therefore, in this section we assume $D=E$
and solve self-consistently the combined system of equations (\ref{8}) and
(\ref{17}). Here we are interested in the dynamics
of the transmission of an electromagnetic pulse incident from vacuum on the
interface with a conductive medium, as well as of the
reflection from this interface. We shall consider low-frequency pulses, in the
sense that their relevant spectral components have
frequencies lower than the collisional frequency. In this situation the Drude
equation may be effectively substituted by Ohm's law. As a matter of fact, we
shall not consider a semi-infinite medium but a finite layer of conductive
material, with its first interface located at $z=0$ and its second interface at
$z=L$. Thus we may write the medium conductivity as $\sigma (z)=\sigma_0$ for
$0<z<L$, and zero otherwise.

In vacuum, at $z<0$, the field can be separated in forward $E_f$ and backward
$E_b$ traveling pulses:
\begin{equation}
E(z,t)=E_f(z-ct)+E_b(z+ct)\, .
\label{20}
\end{equation}
According to the chosen geometry, we identify $E_f$ with the incident pulse, and
$E_b$ with the reflected pulse. The radiation
transmitted through the layer (i.e. at $z>L$) has the form
\begin{equation}
E(z>L,\, t)=E_{tr}(z-ct)\, .
\label{21}
\end{equation}
If the value $\sigma_0$ is small enough, it is possible to solve the wave
equation (\ref{8}) for the
conductive medium by means of a perturbation expansion. In fact, we may expand 
the field
as
\begin{equation}
E=E_0+E_1+\dots ,\quad E_n \sim (\sigma_0 L/c)^n\, .
\label{22}
\end{equation}
In the zeroth order (which corresponds to setting $\sigma_0=0$), the solution is
the forward-traveling pulse
\begin{equation}
E_0=E_f(z-ct)=E_f(\xi )\, ,\quad \xi =z-ct\, ,
\label{23}
\end{equation}
where $E_f(\xi )$ is the shape of the incident pulse. In the first order, the
transmitted field has the same shape as the incident pulse,
but with a reduced amplitude:
\begin{equation}
E_{tr}(\xi )=\left( 1-\frac{2\pi\sigma_0}{c}\, L\right)\, E_f(\xi )\, .
\label{24}
\end{equation}
The shape of the reflected pulse has a less trivial relationship to the incident
pulse:
\begin{equation}
E_b(\eta )=-\frac{\pi\sigma_0}{c}\int_{-\infty}^\eta d\theta\,\left[E_f(-\theta
)-E_f(2L-\theta )\right]\, ,
\label{25}
\end{equation}
where $\eta =z+ct$.
The details of the derivation of Eqs.~(\ref{24}) and (\ref{25}) are provided in
the Appendix.
The two terms in the square brackets of Eq.~(\ref{25})
correspond to two reflections, namely from the first and the second interface,
respectively.
As a result of the integration in Eq.~(\ref{25}), the high-frequency components
which are originally present in the incident pulse are to a certain extent
averaged away from the reflected pulse. In the
case when the thickness of the conductive layer is much smaller than the spatial
extension of the incident pulse, Eq.~(\ref{25})
simplifies to
\begin{equation}
E_b(\eta )=-\frac{2\pi\sigma_0}{c}\, L\, E_f (-\eta )\, .
\label{26}
\end{equation}

The energy fraction that is lost by the transmitted pulse (when compared with
the energy of the incident pulse)
is proportional to the small
parameter $\sigma_0L/c$. In turn, the energy of the reflected pulse is
proportional to the square of the same small parameter.
Therefore, we may conclude that the energy which is absorbed inside the layer is
also proportional to $\sigma_0L/c$.

As it follows from Eq.~(\ref{25}), a pulse which is reflected from a
sufficiently thick layer
(i.e., thicker than the spatial extension of the incident
pulse) may turn out to be much longer than the incident pulse. This reflects the
observation that the Fresnel reflection and
transmission coefficients at the interface between vacuum and the conductive
medium are singular at zero frequency: namely,
the derivatives with respect to frequency of the Fresnel coefficients are
infinite in this limit.
Physically, such singularity would correspond to an infinite delay for
narrowband pulses as their carrier frequency approaches zero.
In the case of broadband electromagnetic pulses, the singularity leads to the
appearance of long tails on their trailing edge.

It proves interesting to complement the above presented analytical results
with specific numerical simulations. Our numerical procedure is based on
the finite-difference time-domain method, which is well suited for solving
initial-value problems for Maxwell's equations.
This permits us to avoid the usual unidirectional
approximation, which is clearly ill-posed for the problem of interest,
simply because the reflection coefficient from the interface between vacuum
and the conductive medium is equal to unity at zero frequency. Additionally, in
our simulations
no slowly varying approximation with respect to time and the spatial
coordinate is used, as we solve the full one-dimensional Maxwell's equations.
Throughout the paper, we choose an incident pulse
which is represented by the sum of the two terms
\begin{eqnarray}
E(z, t=0) & = & \theta (-z)\Big\{
A_0\mbox{sech}\left(\frac{z-z_0}{\Delta z_0}\right)
\label{27}\\
& + & A_1\mbox{sech}\left(\frac{z-z_0}{\Delta z_0}\right)\times\cos\left[(z-z_0)+\phi_0\right]
\Big\}\, .
\nonumber
\end{eqnarray}
The first term in Eq.~(\ref{27})
is a video-pulse, whose spectrum is centered at zero frequency. The second term
is an optical (we shall
refer to it as femtosecond) pulse, whose carrier frequency $\omega_0$ is
relatively far from zero frequency when compared with its spectral
width. Temporal and spatial variables are measured in dimensionless units,
normalized to the carrier
 frequency and the carrier wave number $k_0=\omega_0/c$, respectively. We choose
$\Delta z_0=10$ (i.e.,
 the spatial extensions of both
video- and femtosecond pulses are equal). This choice assures that the two
pulses are well-separated in the spectral domain. Moreover,
with this choice the second pulse contains a relatively small number of optical
oscillations,
so that we can properly call it a femtosecond pulse. The parameter
$z_0=50$ is the initial distance of the peak of the incident pulses from the
interface with the conductive medium.
Such distance is set to be far enough, so that initially no energy
leaks into the medium. To fully assure that initially no energy is contained in
the medium, we multiply the input electric field in Eq.~(\ref{27}) by the step
function $\theta (x)=[1+\mbox{sign}(x)]/2$. $\phi_0$ has the meaning of an
initial phase shift.  $A_0$ and $A_1$ are the amplitudes of the two incident
pulses, and their ratio $A_0/A_1$ is set to $0.2$. In the remainder of the
paper, we will keep these input field parameters unless otherwise specified.

The separation of the incident pulse in vacuum (E=H=B) into the
two above described sub-pulses allows us the simultaneous investigation of two
separate spectral domains: low-frequencies and
high-(or optical) frequencies. In the low-frequency domain, we may follow the
evolution of the (magnetic) area, Eq.~(\ref{4}), which is initially equal to
\begin{equation}
S_B(0)=S_0+S_1\, ,
\label{28}
\end{equation}
where
\begin{equation}
S_0=\pi A_0\Delta z_0
\label{29}
\end{equation}
is the area of the video-pulse, and
\begin{equation}
S_1=\pi A_1\Delta z_0\mbox{sech}(\pi \Delta z_0/2)\cos\phi_0
\label{30}
\end{equation}
is the area of the femtosecond pulse. The latter periodically varies with
$\phi_0$. In our simulations we choose $\phi_0=0$. Since
the incident femtosecond pulse is relatively long, the area of the video-pulse
dominates in Eq.~(\ref{28}).

\begin{figure}
\begin{center}
\includegraphics[scale=1]{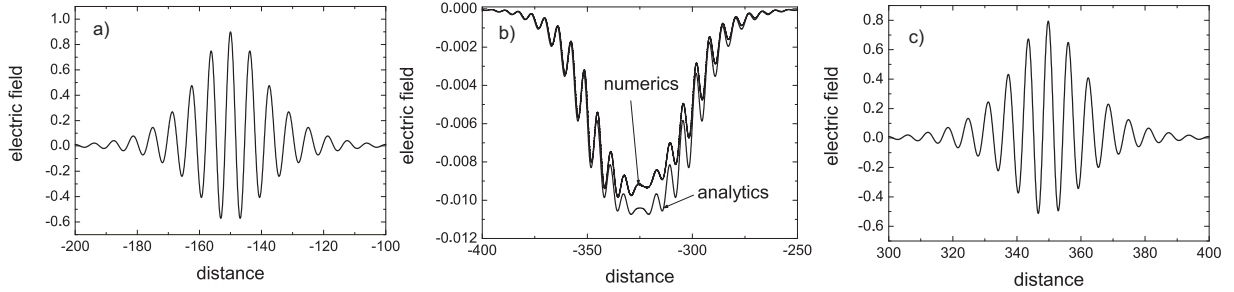}
\end{center}
\caption{Snapshots of the electric field for the incident (a), reflected
(b) and transmitted (c) pulses. Incident pulse given by Eq.~(\ref{27})
with $z_0=150$.}
\label{ris1}
\end{figure}

In order to compare the analytical results with the numerical computations, we
limit ourselves to consider a thin layer of conductive
medium, namely $4\pi\sigma_0=0.01$ and $L=25$, so that the small parameter
$(4\pi\sigma_0/c)L=0.25$. We chose a long enough
simulation time $T$ ($T=500$) in order to allow for almost full reflection from
the boundary. (Recall that we measure time in
units of $\omega_0^{-1}$, $\nu$ and $\sigma_0$ in units of $\omega_0$, and the
length in units of $k_0^{-1}$.) Strictly speaking, simulating a full reflection
would require an infinitely long computation time, owing to the singularity of
the reflection coefficient at zero frequency. As a matter of fact, the tail
following the
pulse has a very low intensity and lasts virtually indefinitely. From the
energetic standpoint, such pulse tail carries a vanishing fraction of energy.
Nevertheless, its contribution to the pulse area
remains substantial, as we shall see later with the calculation of the magnetic
area. The incident, reflected and transmitted pulse shapes are shown in
Fig.~\ref{ris1}: the comparison among the analytically predicted and the
numerically calculated reflected pulses shows a relatively good agreement. These
numerical results are obtained on the basis of the full
Maxwell-Drude model with $\nu =100$ and $\omega_p^2=\omega_0^2$, that is in the
limit case of validity of the Ohm's law.

Figure~\ref{ris1} shows that the pulse which is reflected from the conductive
medium has the form of a video-pulse whose bell-like shape is slightly modulated
by the
optical carrier frequency. Note that the spatial extension of the reflected
pulse is about $1.5$ times longer than the incident pulse.
Therefore the conductive medium acts as a low(high)-pass filter in
reflection(transmission): high-frequency components are
mainly transmitted and somewhat absorbed by the medium, while the reflected
pulse predominantly consists of low-frequency
components. Such filtering properties may be used in practice to discriminate
between low- and high-frequency components of the electromagnetic field. The
numerically computed
transmitted pulse has the same shape as the incident pulse but with a reduced
amplitude, in full agreement with the analytical prediction of Eq.~(\ref{24}):
the respective curves are visually indistinguishable, so we did not show their
comparison in Fig.~\ref{ris1}c. Note that the Fresnel coefficients predict a 
full
reflection of low-frequency components from the conductive medium. However,
such conclusion is only valid for a semi-infinite medium; when dealing
with a thin layer one may still observe an appreciable transmission of
low-frequency components.

Let us turn now to consider the case of a conductive medium layer with a
relatively thicker depth $L$ and higher static conductivity $\sigma_0$, so that
$\sigma_0L/c$ is no longer a small parameter. Under these conditions, the
validity of the analytical perturbation approach is no longer justified. In 
fact,
this situation may be rather described by the approximate diffusion equation
(\ref{10}). Here we proceed with a comparison between the properties of 
solutions of the diffusion equation
(\ref{10}) for the magnetic field, with the numerical solutions of
the Maxwell-Drude system of equations with initial conditions
given as in Eq.~(\ref{27}) with $A_1=0$ (i.e., we only consider an input 
video-pulse).

In order to provide a better illustration of the diffusive
dynamics of the field, we computed the spatial distributions of both electric 
and magnetic fields
inside the conductive medium as time evolved. We found that the electric and 
magnetic field distributions appear to
be qualitatively different. In fact, as a result of the flipping of the phase 
upon reflection from the
boundary, the electric field develops negative regions in the vicinity of the
boundary, which almost completely balance the positive regions situated farther
from the boundary, so that $S_E^{(tr)}\equiv\int_0^\infty dz \, E(z,\, t)\approx 
0$.
As a consequence of this reflection dynamics, the value of the electric field at
the boundary remains appreciably different from zero, and therefore the formula 
(\ref{12_5}) is not
directly applicable. More precisely, although the electric field eventually also 
exhibits a diffusive behavior,
however such behavior is observed for relatively longer (when compared with the 
case of the magnetic field) observation times and also far from the boundary. In 
this respect, we would like to
emphasize that although the diffusion equation (\ref{10}) also correctly 
describes the evolution of
the electric field at all times and everywhere inside the medium, because of its 
peculiar boundary conditions one may not simply apply to the electric field the 
analytically tractable model of a transparent boundary.

\begin{figure}
\begin{center}
\includegraphics[scale=0.7]{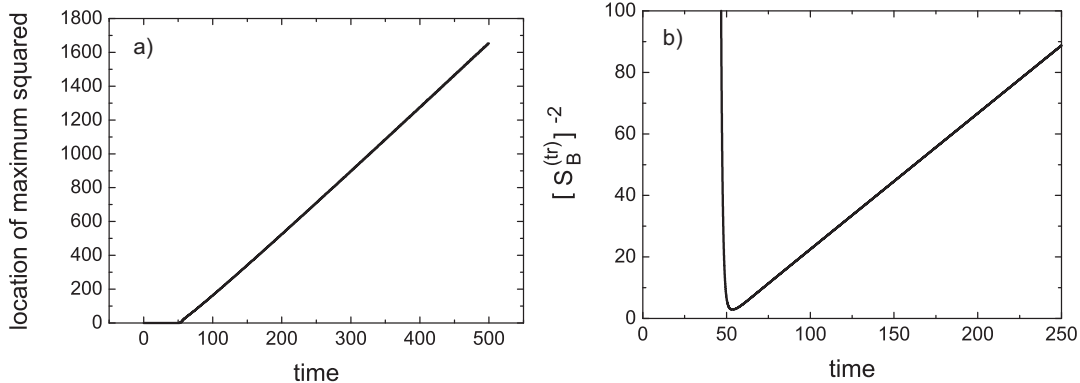}
\end{center}
\caption{a) Location of the square of the maximum of the distribution of the 
magnetic field
inside the medium as function of time; b) the inverse of $[S_0^{(tr)}]^2=[S_B^{(tr)}]^2$ 
as
function of time. Parameters are $\nu =10$, $4\pi\sigma_0=0.5$.}
\label{ris2}
\end{figure}

In contrast, it turns out that the diffusive behavior of the magnetic field is 
better predictable
on the basis of the simple transparent boundary model. Indeed, the reflection of 
the magnetic field is
not accompanied by a flipping of its phase, and as a consequence no negative
field regions appear inside the conductive medium. Moreover, the magnetic field 
quickly approaches zero
at the boundary. This leads to favorable conditions for the application of 
analytical
formulas based on the distribution (\ref{12_5}). First, in our numerical 
simulations we observed that the maximum of
the magnetic field distribution indeed moves with time as $(t-t_0)^{1/2}$. This 
means that the square of this quantity
grows linearly with time, a behavior which is well supported by the plot in 
Fig.~\ref{ris2}a that was
obtained by the numerical integration of the Maxwell-Drude model. The slope of 
the straight
line in Fig.~\ref{ris2}a ($3.77$) is slightly different from the analytically 
predicted value of $4$. This
discrepancy should be attributed to the difference in boundary conditions 
between the real case and the simple analytical model. We also
plot in Fig.~\ref{ris2}b the inverse of the square of the zeroth momentum (i.e. 
of the
magnetic area of the transmitted pulse $S_B^{(tr)}=S_0^{(tr)}$). In support of 
the
analytical predictions that are based on Eq.~(\ref{12_5}), again we observe an 
evolution described by a straight line
immediately after that the incident pulse hits the boundary. From these 
observations we may
conclude that the dynamics of the magnetic field inside the conductive medium is
indeed of a diffusive nature.

\section{The Maxwell-Drude-Bloch model}
Let us include now in our model the presence of bound electronic states. This
formally leads in the emergence of a medium
polarization ${\textbf P}$, which is related to the electrical induction
${\textbf D}$ via the well-known formula:
${\textbf D}={\textbf E}+4\pi{\textbf P}$. According to our model, the resonant
component of the medium consists of a
homogeneous mixture of two types of two-level doping centers, namely the
absorbing (labeled by the index ``p", i.e. passive)
and the amplifying (labelled by index ``a", i.e. active) centers. The total
medium polarization in its scalar form is thus given by
\begin{equation}
P=N_pd_p\rho_{12}^{(p)}+N_ad_a\rho_{12}^{(a)}+{\textrm c.\textrm c.}
\label{31}
\end{equation}
Here $N_{p,a}$ are the concentrations of passive and active two-level doping
centers, $d_{p,a}$ are the (real) dipole matrix elements of transitions between
upper (2) and lower (1) states, and $\rho_{mn}^{(p,a)}={\rho_{nm}^{(p,a)}}^*$
are the elements of the density matrix
describing the resonant atom-like systems. For passive centers, the equations
for the density matrix read as
\begin{eqnarray}
&& \frac{\partial}{\partial t}\rho_{21}^{(p)}=-(i\omega_{21}^{(p)}+\gamma_{21}^{(p)})\rho_{21}^{(p)}
-i\frac{d_pE}{\hbar}\, (\rho_{22}^{(p)}-\rho_{11}^{(p)})\, ,
\label{32}\\
&& \frac{\partial}{\partial t}\rho_{22}^{(p)}=-\gamma_{2}^{(p)}\rho_{22}^{(p)}
-i\frac{d_pE}{\hbar}\, (\rho_{21}^{(p)}-\rho_{12}^{(p)})\, ,
\label{33}\\
&& \frac{\partial}{\partial t}\rho_{11}^{(p)}=\gamma_{2}^{(p)}\rho_{22}^{(p)}
+i\frac{d_pE}{\hbar}\, (\rho_{21}^{(p)}-\rho_{12}^{(p)})\, .
\label{34}
\end{eqnarray}
Whereas for active centers one obtains
\begin{eqnarray}
&& \frac{\partial}{\partial t}\rho_{21}^{(a)}=-(i\omega_{21}^{(a)}+\gamma_{21}^{(a)})\rho_{21}^{(a)}
-i\frac{d_aE}{\hbar}\, (\rho_{22}^{(a)}-\rho_{11}^{(a)})\, ,
\label{35}\\
&& \frac{\partial}{\partial t}\rho_{22}^{(a)}=-\gamma_{2}^{(a)}\rho_{22}^{(a)}
-i\frac{d_aE}{\hbar}\, (\rho_{21}^{(a)}-\rho_{12}^{(a)})+p\, ,
\label{36}\\
&& \frac{\partial}{\partial t}\rho_{11}^{(a)}=\gamma_{2}^{(a)}\rho_{22}^{(a)}-\gamma_1^{(a)}\rho_{11}^{(a)}
+i\frac{d_aE}{\hbar}\, (\rho_{21}^{(a)}-\rho_{12}^{(a)})\, .
\label{37}
\end{eqnarray}
For simplicity, let us limit ourselves to consider atom-like doping centers with
homogeneous broadening.
In the above equations, $\omega_{21}^{(a,p)}$ are the transition frequencies,
which are taken to be equal to each other,
as well as to the carrier frequency $\omega_0$ of the incident femtosecond
pulse;
$\gamma_{21}^{(a,p)}$ are the polarization relaxation rates; $\gamma_2^{(a,p)}$
are the upper state population relaxation rates;
$\gamma_1^{(a)}$ is population relaxation rate for the lower state of the active
centers; and finally $p$ is the pump rate.
As it follows from the above presented
model equations, for passive centers one obtains a closed system where the lower
state is the ground state. In contrast, the active
two-level doping centers form an open system, since we supposed that there is a
decay mechanism out of the two-level configuration, as well as pumping from some
auxiliary
upper level. Therefore we may only write the conservation law
$\rho_{11}^{(p)}+\rho_{22}^{(p)}=1$ for passive centers; no such
conservation holds the for the active doping two-level centers.

In our numerical simulations we used the following set of parameters: the ratio
of dipole moments $\sqrt{\mu}=d_p/d_a$ was $1.5$,
the relaxation rates were $\gamma_1^{(a)}=0.025$, $\gamma_2^{(a)}=0.005$, 
$\gamma_2^{(p)}=0.006$,
$\gamma_{21}^{(p)}=0.0025$,
$\gamma_{21}^{(a)}=0.015$, the pump rate was $p=0.004$. All of these quantities
are expressed in units of the frequency $\omega_0$, so that the coupling
constants between the
field (or dimensionless Rabi frequency $\Omega =2d_pE/\hbar\omega_{21}^{(p)}$)
and the polarizations induced by the passive and active
centers read as $\beta =4\pi N_pd_p^2/\hbar\omega_{21}^{(p)}=0.1$ and $\kappa
=4\pi N_ad_pd_a/\hbar\omega_{21}^{(a)}=0.02$.

\begin{figure}
\begin{center}
\includegraphics[scale=1]{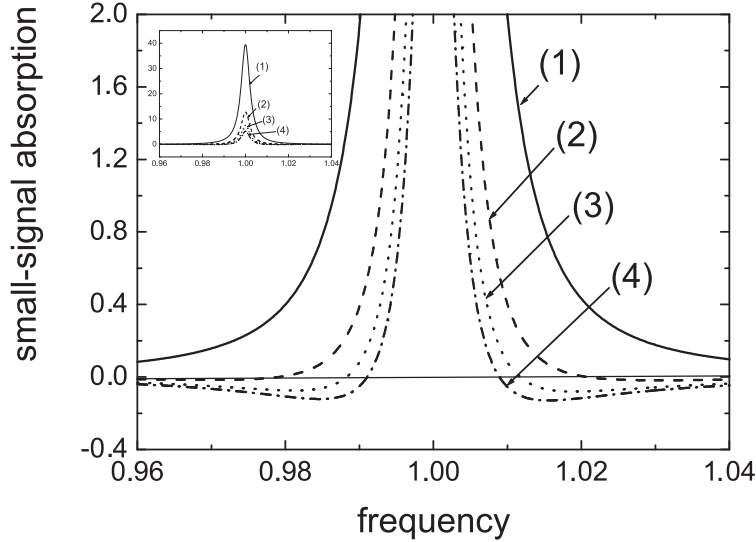}
\end{center}
\caption{Small-signal absorption coefficient $\alpha (\omega )$ defined by
Eq.~(\ref{39}) for four concentration levels of the passive
centers $\beta$: (1) $0.1$; (2) $0.033$; (3) $0.014$; (4) $0.011$. Other
parameters are as introduced in the text, moreover
$\nu =10$ and $4\pi\sigma_0=0.005$. The curve (1) is everywhere positive,
corresponding to the stability of the trivial solution. Full
curves are shown in the inset.}
\label{ris3}
\end{figure}

From now on, we are going to solve in a self-consistent manner by the
finite-difference time-domain method the coupled system
including the wave equation (\ref{8}), the Drude equation (\ref{17}), and
Bloch equations (\ref{32})-(\ref{37}). Before describing the full numerical
solutions of this Maxwell-Drude-Bloch (MDB) model, let us
present the analytical stability analysis of its trivial solution ${\textbf
E}=0$. Such analysis was described in
a general form at the end of Section II. With a dispersion relation as in
Eq.~(\ref{16}), we need to specify two ingredients, namely the
frequency-dependent dielectric permittivity $\varepsilon (\omega )$ and the
conductivity $\sigma (\omega )$. The latter is given in
Eq.~(\ref{19}), whereas the former can be found from the linearized version of
the density matrix equations (\ref{32})-(\ref{37}). One obtains
\begin{equation}
\varepsilon (\omega )=1-\beta N_{eq}^{(p)}\frac{\omega +i\gamma_{21}^{(p)}}{\omega^2+{\gamma_{21}^{(p)}}^2}
-\kappa\sqrt{\mu}N_{eq}^{(a)}\frac{\omega +i\gamma_{21}^{(a)}}{\omega^2+{\gamma_{21}^{(a)}}^2}\,
.
\label{38}
\end{equation}
Here $N_{eq}^{(p)}$ and $N_{eq}^{(a)}$ are the normalized equilibrium population
differences
of passive and active centers in the absence of electric
field. For the above chosen parameters, one obtains $N_{eq}^{(p)}=-1$ and
$N_{eq}^{(a)}=0.64$. By inserting Eqs.~(\ref{19}) and (\ref{38}) in the
dispersion relation Eq.~(\ref{16}), we may describe the propagation of a weak
radiation in the conductive medium.
For the trivial solution ${\textbf E}=0$ to be
stable, radiation at all frequencies should decay upon propagation, so that the
overall small-signal absorption coefficient
\begin{equation}
\alpha (\omega )=\mbox{sign \{Re}[k(\omega )]\}\, \mbox{Im}[k(\omega )]
\label{39}
\end{equation}
must be non-negative for any positive value of $\omega$. A graphical
representation of the frequency dependence of the
small-gain absorption coefficient is given in
Fig.~\ref{ris3}. In this figure it is shown that, among the four curves shown
here which correspond to different concentrations of the two-level absorbers,
the stability condition is only satisfied for curve (1). Therefore, we used the
corresponding concentration $\beta=0.1$ in our subsequent numerical simulations
of the dissipative soliton generation. As it is usual for nonlinear dissipative
systems, nonlinear gain may exceed nonlinear absorption, even though in the
linear limit absorption prevails. The
switching between the two regimes occurs due to the effect of a nonlinear (in
our case, also coherent) bleaching of the absorption.

\begin{figure}
\begin{center}
\includegraphics[scale=0.75]{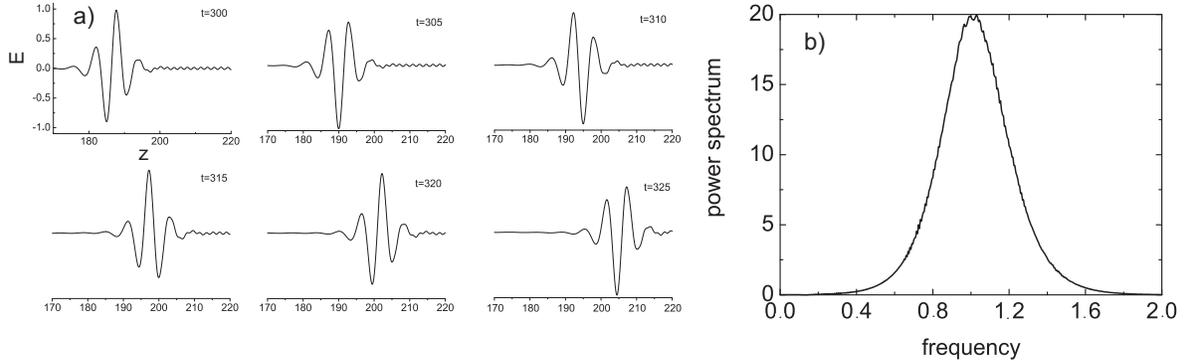}
\end{center}
\caption{a) Profiles of the electric field (normalized Rabi frequency) of a
few-cycle MDB dissipative soliton at different instants of time inside the
conductive medium. The total tracking time covers approximately one period of
the quasi-periodic motion of the soliton profile.
b) The soliton spectrum. Specific parameters are $\nu =0.1$,
$4\pi\sigma_0=0.005$.}
\label{ris4}
\end{figure}

Given the above considerations, we are now ready to turn our attention to the 
full
nonlinear problem.
As it was previously discussed in \cite{VRS_06}, in a zero conductivity medium
containing a purely resonant ensemble of amplifying and absorbing atom-like
centers, few-cycle stationary pulses
are metastable. In other words, for few-cycle pulses that propagate in such
media there is a critical value of energy, such that pulses with lower (higher)
energy eventually disperse (collapse). However, as we shall see in the
following, in the presence of a nonzero and wideband conductivity as in the
present MDB model, the pulse collapse
effect may be suppressed. Indeed, we found that such stabilization of the pulse
energy may take place for a wide range of variation of the
collisional frequency $\nu$. We numerically checked the occurrence of the
dissipative soliton energy stabilization process in the MDB model for three
representative regimes, namely for $\nu =0.1$, $1$ and $10$ (in units of the
carrier angular frequency $\omega_0$).
As we are going to demonstrate a little later, the regime with $\nu\ll\omega_0$
turns out to be quantitatively closer to realistic semiconductor parameters. An
example of numerically generated dissipative soliton within the MDB framework is
illustrated in Fig.~\ref{ris4}. The duration of this MDB soliton is less than
two optical periods, and its
spectrum (which is also shown in Fig.~\ref{ris4}) is extremely wide (coherent
supercontinuum). In fact the MDB soliton spectral width is comparable to the
value of the
transition frequency itself, and its spectrum is exactly centered at the
transition frequency, in contrast with video-solitons
(that were observed in a passive two-level scheme in Ref.~\cite{BA_71}) as well
as the solitons that are formed in active-passive three-level ensembles of
doping centers (see Refs.~\cite{VRS_06,RSV_08}). We did not observe generation
of video-solitons in the MDB model. Physically, such property is associated with
the
high-pass filtering property of the conductive medium. Note that a similar
situation was also observed in the
dissipative soliton generation scheme which was presented in Ref.~\cite{VRS_09},
where active and passive centers were embedded into a dielectric matrix that was
characterized
by an infra-red absorption band.

The quasi-periodicity of the temporal profiles shown in Fig.~\ref{ris4} is
related to the fact that the MDB soliton carrier frequency is distant from zero.
If the soliton was propagating in vacuum, all of the snapshots which are shown
in the figure would look exactly the same. However, whenever the soliton
propagates in the conductive medium with both active and passive two-level
doping centers,
the soliton spatial profile varies in time. This is analogous to the case of the
well-known self-induced transparency theory, where the propagation speed of the
soliton envelope is different from the carrier phase velocity. A similar
situation occurs in our case: we followed the zero-crossing of the field as it
traverses through the (imaginary) pulse envelope, and we concluded with $1$\%
accuracy that their speed is equal to $c$, independently upon the precise
location of a zero inside the envelope.
In parallel, we also followed the speed of the motion of the soliton peak: for
this particular case, we obtained a group speed of $0.69c$.  Therefore we may
indeed conclude that the carrier moves much faster than the (imaginary)
pulse envelope.

\begin{figure}
\begin{center}
\includegraphics[scale=0.75]{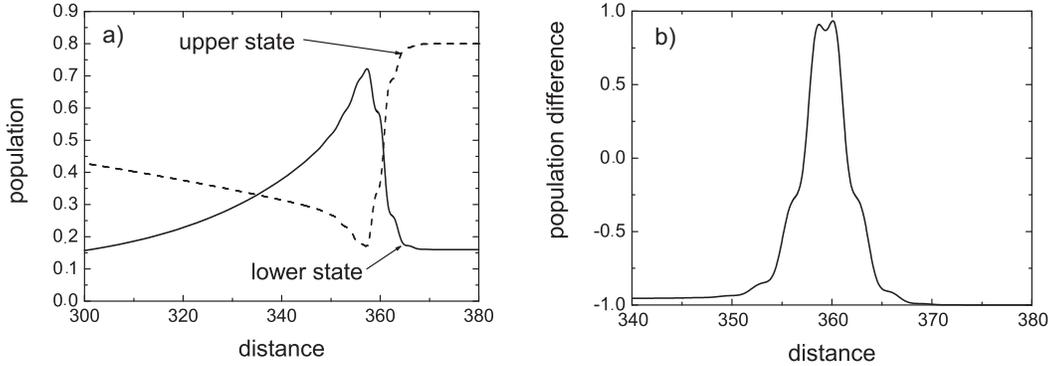}
\end{center}
\caption{a) Spatial dynamics of the populations of the upper and lower states
in the amplifying doping centers;
b) spatial dynamics of the population difference in the absorbing centers.
The parameters are the same as in Fig.~\ref{ris4}, and the propagation
time is $500$. Note the almost full inversion occurring at $z=357$ in the
amplifiers.}
\label{ris5}
\end{figure}

Note that the weak ripples that modulate the spatial profiles (and
correspondingly the spectrum) of the soliton pulse in Fig.~\ref{ris4} result
from the initial transient process that leads to the formation, from the
incident pulse, of a dissipative soliton inside the conductive medium. The input
pulse that is incident from vacuum was set according to the expression in
Eq.~(\ref{27})
with $A_1=0.75$. The profile of the soliton electric field which is illustrated
in Fig.~\ref{ris4} is associated with corresponding spatial evolutions for the
populations in both active and absorbing doping centers, as it is shown in
Fig.~\ref{ris5}. In particular, the evolution of the population difference among
the two levels of the absorbing centers, as it is illustrated in  
Fig.~\ref{ris5}b,
reveals that almost complete inversion is achieved in the
middle of the MDB soliton pulse. Whereas after the pulse the atomic-like systems
return back to the ground state (except for a small tail which is the result of
relaxation processes). Such
behavior resembles the complete cycling of the population which is initiated by
a $2\pi$-pulse of self-induced transparency in the theory of long optical
pulses. In parallel, for the active centers the exchange among the populations
of the upper and lower levels that is observed in the left side of
Fig.~\ref{ris5} is analogous to the behavior of $\pi$-pulses, which have the
property to invert the populations in a coherent amplifier. For relatively
longer distances (not fully shown in the scale of the figure), the two-level
populations of the active centers slowly restore their initial values as a
result of the combined action of the relaxation and pumping processes.

\section{MDB soliton formation}
\begin{figure}
\begin{center}
\includegraphics[scale=0.75]{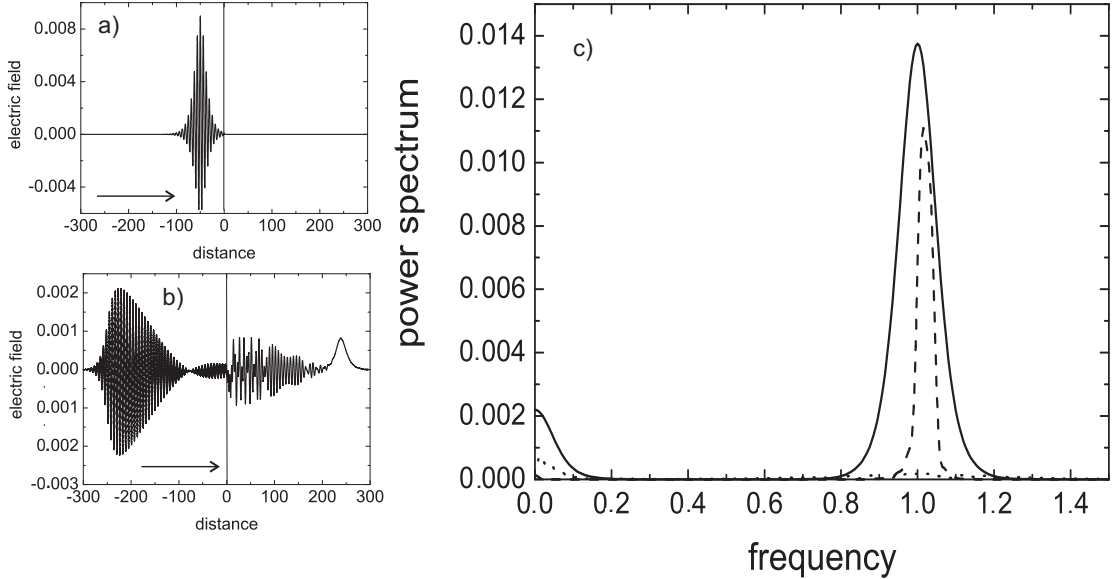}
\end{center}
\caption{a) The incident pulse at $t=0$, and b) the pulse at $t=300$
in the linear regime of interaction. The vertical
line indicates the boundary, arrows --- the direction of propagation of the
incident pulse. c) Spectra of the incident (solid), reflected
(dashed) and transmitted (dotted) pulses. The parameters are:
$\nu =10$, $4\pi\sigma_0=0.005$.}
\label{ris6}
\end{figure}

The generation of MDB few-cycle solitons from an incident femtosecond pulse that
we discussed in the previous section exhibits a threshold behavior, much in the
same way as for other types of dissipative solitons. In other words, for a given
set of medium parameters, a specific energy threshold exists for an incident
pulse of a given shape: the dissipative soliton can only be formed whenever the
incident pulse energy is larger than a specific threshold value. The existence
of such an energy threshold is a direct consequence of the above discussed
stability property of the trivial solution ${\textbf E}=0$. For the specific
incident pulse shape that is given by
Eq.~(\ref{27}), we may define a threshold value for the amplitude of the
femtosecond pulse. Whenever
$\nu =0.1$ and $4\pi\sigma_0=0.005$, such amplitude threshold is equal to
$A_1=0.35$.
Signals with smaller amplitudes are partly absorbed and partly reflected by the
conductive medium.
An example of pulse evolution in the case of an incident femtosecond pulse whose
amplitude is below the threshold is
shown in Fig.~\ref{ris6}: we may notice from this figure that a remarkably large
fraction of the incident
optical pulse energy is back reflected at $t=300$.
Such a strong reflection is due to
resonantly-enhanced index of refraction of the two-level dopants of the medium.
On the other hand, in the case of Fig.~\ref{ris6}
the value of the conductivity and the
interaction time are not large enough in order to cause a substantial reflection
of the low-frequency electric field components.
As a result,
these components are virtually absent from
the reflected pulse spectrum, which basically only includes high-frequency
components, albeit slightly blue-shifted with respect to the resonance
frequency. Figure~\ref{ris6} also shows that the resonant dispersion strongly
affects the temporal shape of the reflected pulse.
On the other hand, the transmitted pulse energy is relatively reduced with
respect to the reflected counterpart. Indeed, Fig.~\ref{ris6}
shows that the transmitted pulse is formed by the undistorted remainder of the
incident video-pulse moving with speed close to $c$, with the
addition of an high-frequency optical pulse whose irregular
envelope travels with a group speed lower than $c$.

\begin{figure}
\begin{center}
\includegraphics[scale=0.75]{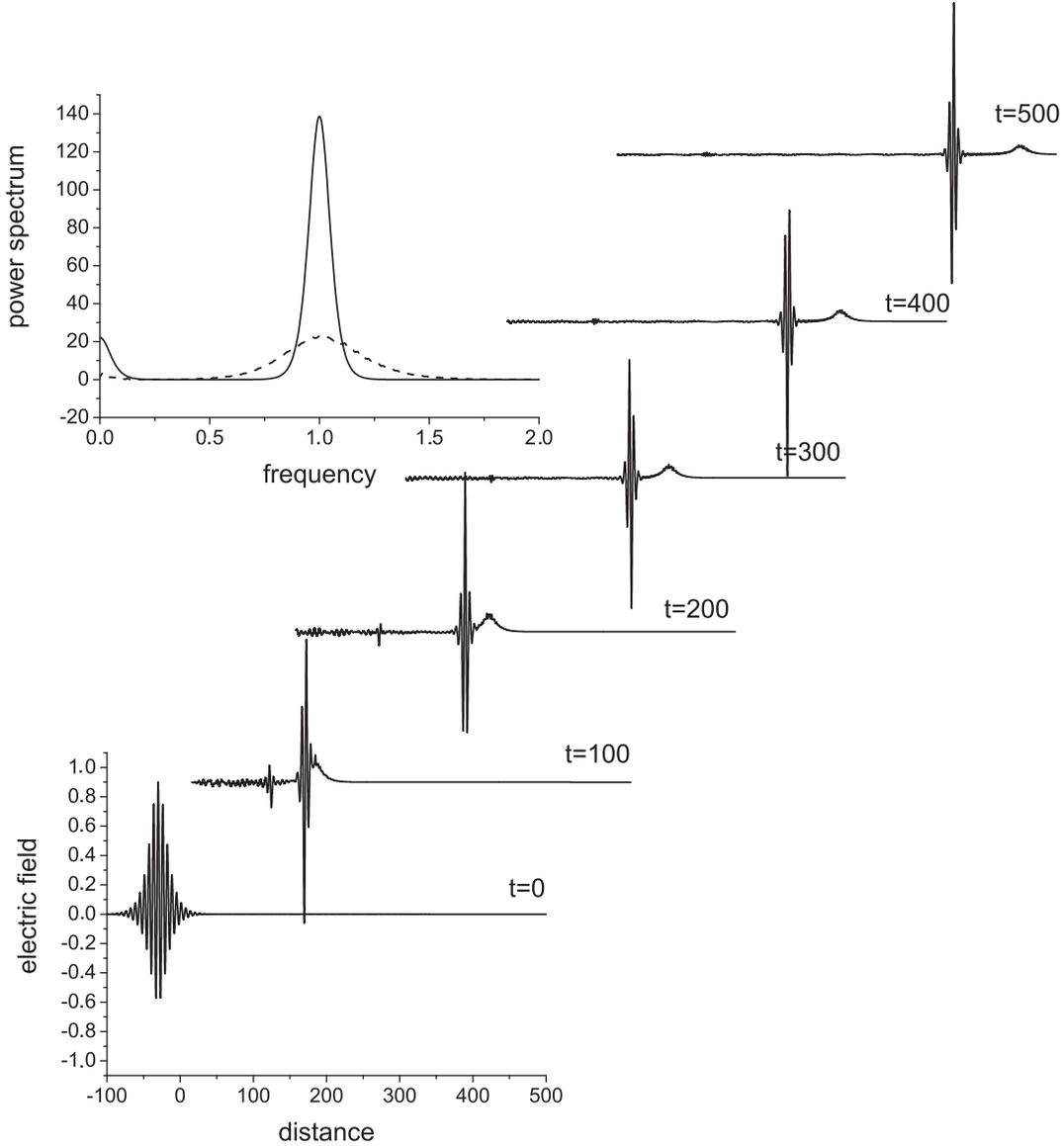}
\end{center}
\caption{Evolution plots of the initial combination of a femtosecond pulse with
$A_1=0.75$ and a video-pulse with $A_0=0.15$, showing
soliton formation in a medium with parameters $\nu =10$ and $4\pi\sigma_0=0.005$.
We also show
the spectra of the incident (solid) and final (dashed) pulses.}
\label{ris7}
\end{figure}

On the other hand, the evolution of an incident femtosecond pulse with an
above-threshold amplitude results in
the formation of a dissipative MDB soliton as discussed in the previous section
(see also Fig.~\ref{4}). The dynamics of such evolution is illustrated in
Fig.~\ref{ris7}:
for this particular
realization, the transient process leading to
soliton formation from the incident pulse takes a few hundreds of dimensionless
time units.
The amplitude of the incident pulse, $A_1=0.75$,
was chosen to be sufficiently large, so that in the vicinity of the boundary the
optical frequency part of the transmitted pulse
is separated into two sub-pulses that move with
drastically different speeds. The fast-moving optical sub-pulse gradually
transforms into a dissipative soliton; whereas its
slowly-moving counterpart
remains relatively close to the boundary, until eventually it dissipates away
all of its energy. The gain which is
provided by the active component of the
medium is strongly depleted by the leading pulse,
so that any subsequent pulse does not experience enough gain to compensate for
its dissipative
energy decay. In this sense, we may anticipate that the medium does not support
the formation of multiple MDB solitons in the case of single pulse excitation: 
at least we
did not detect such patterns in our simulations.
Upon propagation in the conductive medium, the optical pulse energy in
Fig.~\ref{ris7} slightly decreases, until a stationary
value is reached. On the other hand, for relatively smaller values of the
initial amplitude $A_1$ of the incident
femtosecond pulse (still, with $A_1$ greater than $0.35$),
we observed a growth of the pulse energy from its initial value up to a common
asymptotic value.
In contrast with the previously discussed linear regime of
interaction with the medium, in the nonlinear case the fraction of incident
optical
pulse energy which is reflected from the boundary remains relatively small. In
other words, we may say that at high power levels
the boundary is virtually transparent, owing to the nonlinear saturation of the
resonance of the two-level doping centers.

Figure~\ref{ris7} also shows that the dissipative soliton generation process is
accompanied by the propagation of a transient faster video pulse, or precursor,
which gradually separates from the optical pulse as the field penetrates from
the interface into the conductive medium. Indeed, such precursor results from
the video pulse component of the incident field of Eq.~(\ref{27}), and it
propagates with speed close to $c$. As a matter of fact, at low frequencies the 
electric
field is out of resonance with the two-level doping centers, hence it is not
subject to the slowing down which caused by the
resonance mechanism. Nevertheless, the transmitted video pulse decays
exponentially in time, as it transfers its energy to free electrons via the
Drude conductivity.
In summary, we found that the ultimate shape of the few cycle optical soliton is
virtually independent upon the precise details of the input field excitation
process. In
particular, the MDB soliton shape does not depend on the presence or absence of
input video-pulse: exactly the same few-cycle optical
soliton could be generated with
$A_0=0$.  Moreover, the profiles of the MDB soliton electric and magnetic fields
are almost indistinguishable. In Fig.~\ref{7} we also compare the spectra of the
incident and the
transmitted pulses: here the spectrum of the transmitted pulse coincides with
the soliton spectrum
that was earlier illustrated in Fig.~\ref{4}. Note the substantial spectral
broadening of soliton spectrum with respect to the spectrum of the incident
femtosecond pulse, thanks to the effect of temporal compression that has been
induced by the medium.

\begin{figure}
\begin{center}
\includegraphics[scale=0.75]{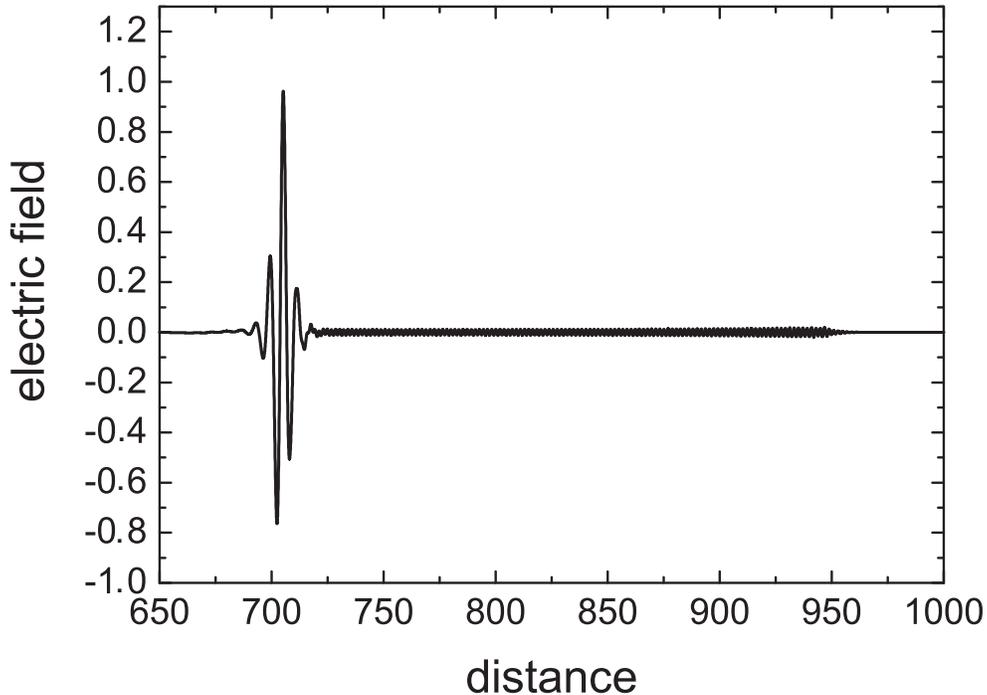}
\end{center}
\caption{The shape of the pulse after $1000$ time units of propagation in the
conductive medium with $\nu =0.1$ and
$4\pi\sigma_0=0.005$. Note that the precursor is spread out over large distance,
while the soliton preserves the shape found in
other simulations.}
\label{ris8}
\end{figure}

We also performed additional simulations with $\nu =1$ and $\nu =0.1$,
while keeping unchanged the value of the conductivity at the resonance
frequency. In
all of these cases we obtained evolution plots for the electric field which do
not differ much from that shown in Fig.~\ref{ris7}, so that it would be
redundant to present them all.
Instead, we plot in Fig.~\ref{ris8} the final temporal profile of the MDB
soliton pulse after $1000$ time units of propagation in the conductive medium
with
$\nu =0.1$ and $4\pi\sigma_0=0.005$. The only visible difference with respect to
the final MDB soliton profile that was shown in Fig.~\ref{ris7} is the
relatively much larger spreading of the precursor: instead of the compact,
bell-type shaped video pulse of Fig.~\ref{ris7}, we obtained a low-amplitude
precursor with a modulation
frequency $3.5$ times larger than the resonance frequency. In this case the
breakdown of the input video-pulse may be attributed to the
prevalence of the derivative term over the collisional term in the Drude
equation (\ref{17}).

\begin{figure}
\begin{center}
\includegraphics[scale=0.75]{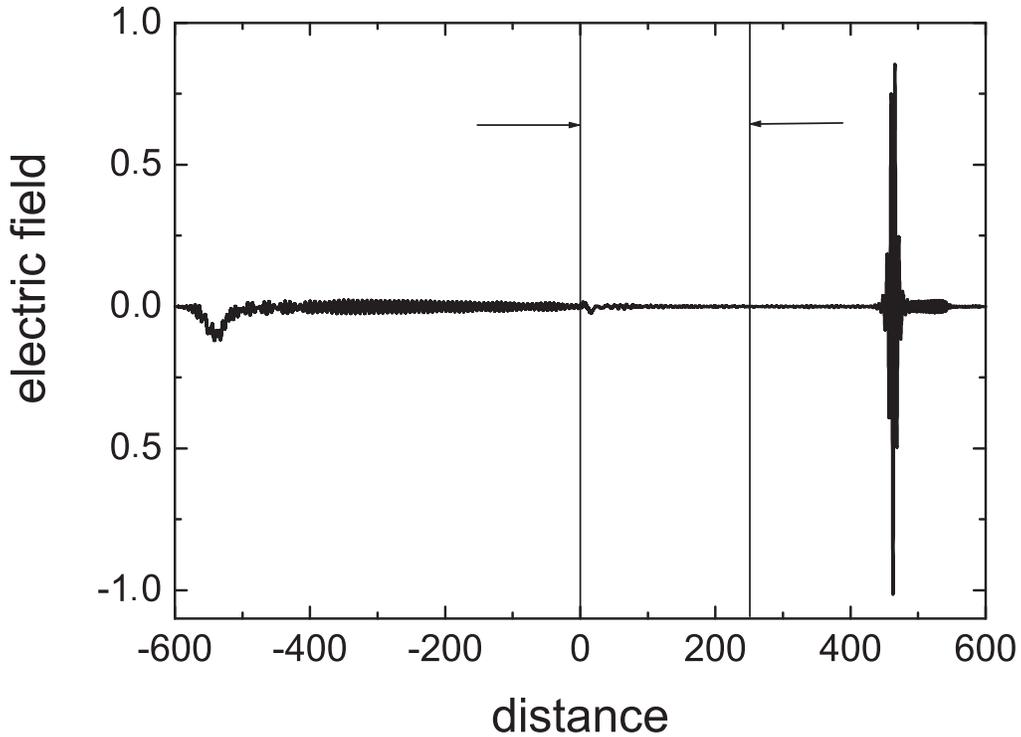}
\end{center}
\caption{Pulse from a finite layer of the medium: a soliton and a precursor are
formed. The parameters are $\nu =0.1$ and $4\pi\sigma_0=0.005$. Boundaries are
indicated by thin vertical lines.}
\label{ris9}
\end{figure}

Before concluding the description of the evolution of the electromagnetic field
within the framework of the Maxwell-Drude-Bloch model,
we show in Fig.~\ref{ris9} the case of a pulse passing through a thick but
finite layer of the conductive medium.
The inclusion of a second
(or output) interface brings our model closer to a real experiment. The layer
was set to be thick enough ($z=250$) in order to allow for completing the
formation of the dissipative MDB soliton from the incident pulse.
The plot in Fig.~\ref{ris9} aims at demonstrating the absence of dramatic
changes to the soliton
shape when passing from the medium through the boundary to vacuum. However in
this case the medium thickness was not enough to
fully dissipate the incident video-pulse. As a result, a precursor still
accompanies the MDB soliton.
Note that the presence of such a precursor is somewhat
artificial. In fact, we included in the incident pulse a video-pulse
component in order to be able of following the evolution of low-frequency
radiation in the conductive medium. Clearly in practical experimental conditions
an incident femtosecond pulse is sufficient for the excitation
of a MDB soliton: in this case no precursor would be observed, apart from
possible radiation initiated in the transient process.

\begin{figure}
\begin{center}
\includegraphics[scale=0.75]{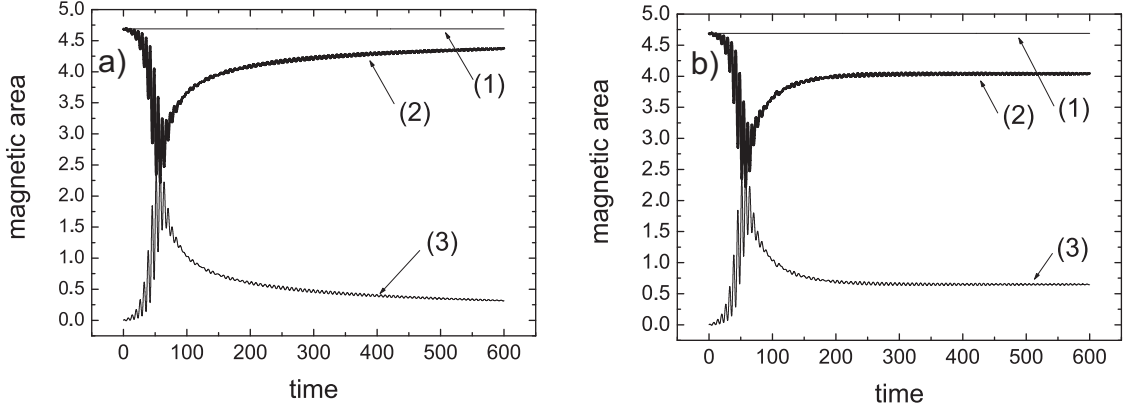}
\end{center}
\caption{Total magnetic area (1), the magnetic area of the reflected (2) and
transmitted (3) pulses for thick, $L=250$, (a) and thin,
$L=25$, (b) layers. Parameters are $\nu =0.1$ and $4\pi\sigma_0=0.005$.}
\label{ris10}
\end{figure}

Finally, let us turn our attention to the discussion of the dynamics of the
evolution of low-frequency pulse components,
or video-pulse.
The conductive component of the medium plays here the dominant role,
whereas the  resonant atom-like systems have a negligible influence.
As we shall see, the notion of magnetic area lies at the heart of this
discussion. Let us consider the total magnetic area as defined according to
Eq.~(4), and the
partial areas
\begin{equation}
S_B^{(tr)}(t)=\int_{0}^\infty dz\, B(z,t)\quad\mbox{and}\quad
S_B^{(ref)}(t)=\int_{-\infty}^0 dz\, B(z,t)
\label{40}
\end{equation}
for the transmitted and reflected pulses, respectively. Clearly, the sum of the
last two areas yields the total magnetic area. Let us recall here that the area
of the magnetic field equals its zero-frequency Fourier component.

Let us examine the two limit cases of a pulse crossing either a thick or a thin
layer of conductive medium. Figure~\ref{ris10} shows the corresponding time
evolutions of the magnetic area. The total magnetic area is a conserved
quantity in both cases, a conclusion that follows from the
analytical formula (\ref{5}) and which is well confirmed by the
computer-generated plots of Fig.~\ref{ris10}. As it can be seen from
Fig.~\ref{ris10}, in the case of a thin medium the two partial magnetic areas
reach their corresponding asymptotic values in a relatively short time. On the
other hand, for a thick layer the magnetic areas of the reflected and the
transmitted pulses do not reach their asymptotic values within the total time of
observation. Such slow time evolution of the magnetic area reflects the
continuous creation of evanescent waves inside the medium. Such evanescent field
is associated with the nearly unitary reflection coefficient at zero
frequencies.
As a result, one obtains a relatively long decay time of the partial magnetic
areas.
It is remarkable that the evanescent field, which appears as a long,
non-oscillating and slowly decaying tail
that follows the main pulse, is not even visible on the scale of previous plots
such as Fig.~\ref{ris9}.
The reason is that this tail is of vanishing intensity, thus it contributes
a relatively small fraction to the total energy of the pulse.
Nevertheless, the electric area of the tail may remain quite substantial, as it
is proportional to the tail
amplitude multiplied by its long spatial extension.
wave.

In the case of a thin layer of the medium,
Fig.~\ref{ris10} shows that the asymptotic values of the partial magnetic areas
are readily achievable within a relatively short observation time. This
corresponds to the fact that here the reflection coefficient is noticeably
different from unity at zero frequencies (frustrated total internal
reflection). In other words, the evanescent wave simply tunnels through the thin
layer of conductive material, and carries the energy of
low-frequency components through the layer. In summary, the analysis of the
temporal dynamics of the magnetic area permits us to investigate the details of
the
propagation of evanescent waves through a finite conductive medium.

\section{Discussion}
In this work we studied the linear and nonlinear dynamics of the propagation of
broadband radiation which is
incident on the interface with a conductive medium, doped with
active and passive two-level atom-like systems. To this end, we introduced the
Maxwell-Drude-Bloch model, which
appears as a promising approach for the description of actual devices in modern
nonlinear optics. In fact, the MDB model allows for the consideration of the
full Lorentz model
of matter, which accounts for the interaction of light with both free as well as
bound electrons.
We anticipate that the dynamics of free charges may become particularly
important when describing doped semiconductor materials. An appropriate
generalization of the model which was presented here will include of the
nonlinear response of the electric current, which is typical of photovoltaic
phenomena.

We have shown that, within the framework of the Maxwell-Drude-Bloch model,
stable few-cycle dissipative solitons can be formed.
A remarkable feature of the present scheme is that these MDB solitons may be
excited by a relatively long (albeit rather powerful) femtosecond pulse which is
incident from vacuum onto the boundary with the conductive medium, doped with
passive and active atom-like systems. It is important to note that the process
of pulse compression leading to soliton formation is not energy-consuming: quite
to the opposite, soliton generation may even be accompanied by a power increase,
thanks to the presence of gain in the medium.
The dissipative solitons can be only excited by rather powerful pulses, with
energies exceeding certain threshold (hard excitation).
These few-cycle MDB solitons are characterized by a wide spectrum (coherent
supercontinuum), whose carrier is centered at the resonance frequency. On the
other hand, given low transmissivity of the conductive medium at low
frequencies, we do not expect the formation of video-solitons by the present
scheme.

The process of formation of few-cycle solitons involves the rather natural
assumption that the initial pulse is incident from vacuum
on the boundary with the conductive doped medium. In this case, one must
properly take into account the
high reflectivity of such a medium at
low frequencies. As a matter of fact, the medium is totally reflective at zero
frequency only, and in the case of both a semi-infinite medium and an
indefinitely long
interaction time. On the other hand, in the case of propagation through a thin
layer of the medium, one only obtains a partial reflection of low frequency
waves, owing to the tunneling of evanescent waves.
As a consequence, in this case instead of total reflection one observes a long
and low-intensity tail of radiation that follows
the transmitted main optical pulse. Even though such a tail carries a relatively
small fraction of energy, its area may remain quite substantial. In order to
characterize these low-frequency tails, we introduced the notion of temporal
magnetic area, see Eq.~\ref{4}. For this quantity we
found a conservation law in the form of Eq.~(\ref{5}), which holds well beyond
the framework of the Maxwell-Drude-Bloch model, that is for virtually any pulsed
electromagnetic radiation that propagates in a homogeneous or inhomogeneous
medium with free and bound
charges and dissipation. We also found it quite instructive to separate the
total magnetic area in two parts -- namely, the transmitted and reflected areas,
as these quantities characterize the low-frequency tails of the transmitted and
reflected pulses, respectively.

For low-frequency pulses, we have derived a parabolic equation (\ref{10}), that
describes the process of diffusion of the
electromagnetic field in a highly conductive medium. Though the parabolic 
equation
has been known in the electromagnetics of quasi-stationary fields, 
\cite{LL,Jackson},
here we reveal its new application to video-pulses and demonstrate the 
propagation
dynamics of these pulses and related evolution of their magnetic area. This 
equation
may be also useful in studies involving the propagation of THz radiation in high
conductivity materials.

In the end, let us present some numerical estimates for the effective values of
the conductivity and the collisional frequency. The latter can be
found if the mobility $\mu =e/\nu m_{\mbox{eff}}$ and the effective mass
$m_{\mbox{eff}}$ of the carriers are known. For GaAs
we get $\mu =0.85$~m$^2$/$(V\cdot s)$ and $m_{\mbox{eff}}=0.067m_e$, so that
$\nu\approx 3\cdot 10^{12}$~Hz; for AlAs:
$\mu =0.028$~m$^2$/$(V\cdot s)$, $m_{\mbox{eff}}=0.1m_e$ and $\nu\approx
6\cdot 10^{13}$~Hz, \cite{semi1,semi2}. Both values of the collisional frequency 
$\nu$ are much smaller than the optical frequency 
($\omega_0\sim 5\cdot 10^{14}$~Hz), which
corresponds to the limit case $\nu\ll\omega_0$. For
infrared radiation one may obtain an intermediate situation $\nu\simeq\omega_0$,
whereas for THz radiation this inequality is reversed,
namely $\omega_0\ll\nu$. In all of these cases we observed the stable generation
of dissipative MDB solitons.

Another important estimate involves the value of the static conductivity, which
is usually measured in SI units. Therefore we rewrite the expression for the
quantity $4\pi\sigma_0$ (which is normalized with respect to the optical
frequency $\omega_0$) that we have used throughout our simulations, as
\begin{equation}
4\pi\sigma_0=\frac{c^2\mu_0}{\omega_0}\,\frac{1}{2}N_e e\mu
\label{41}
\end{equation}
where we used Eq.~(\ref{18}) expressed in CGSE units, and rewritten here in 
terms
of the mobility parameter;
$\mu_0=1.257\cdot 10^{-6}$~H/m is the magnetic permeability of the vacuum. For
$\mu =0.85$~m$^2$/$(V\cdot s)$ and
$\omega_0\sim 5\cdot 10^{14}$~Hz, we get the estimate
\begin{equation}
4\pi\sigma_0=1.3\cdot 10^{-23} N_e(\mbox{m}^{-3})\, .
\label{42}
\end{equation}
For the value of $4\pi\sigma_0=0.005$ that we used in the above simulations for
demonstrating the
generation of dissipative solitons, the concentration of
free carriers is
$N_e=3\cdot 10^{14}$~cm$^{-3}$ (this is approximately equal to the concentration
of dopants in the host semiconductor material). Such value of dopant
concentration is rather modest. Clearly for higher values of the concentration
the conductivity
would be even more significant.

In order to implement the two-level gain and absorbing doping centers,
we may suggest the use quantum dots, i.e. the same medium as in the scheme
with three-level atomic-like doping centers that was proposed in
Ref.~\cite{VRS_09}.
However in the present situation the quantum dots would not be embedded in
quartz as in Ref.~\cite{VRS_09}, but in a
semiconductor environment. Due to their huge dipole moments, the use of quantum
dots may permit a decrease by several orders of magnitude of the peak
intensities of the electromagnetic radiation, thereby avoiding the occurrence of
a nonlinear response or even
breakdown of the host matrix. Some associated
numerical estimates can be found in Ref.~\cite{VRS_09}. Note that for the
specific values, that we used in our
simulations, of quantum dot concentrations of the order of $10^{18}$~cm$^{-3}$,
the thickness of the medium which is necessary for
the development of a dissipative soliton is less than one millimeter. These 
rather
high concentration levels, as well as the relatively fast decay constants into
the asymptotic soliton solution are not crucial to our scheme, and merely served
for permitting a fast numerical convergence of the initial pulse into the
soliton. Therefore the doping concentration
could be reduced by an order of magnitude without preventing the applicability 
of
our scheme, but possibly at the
expense of device miniaturization.

The broadband losses that are introduced by the conductivity represent the main
stabilization mechanism of the dissipative solitons.
We anticipate that in a laser where the active two-level systems play the role
of a gain medium, passive two-level absorbers would play the role of a
mode-locker in a passive mode-locking configuration, whereas the role of the
broadband losses may be taken by the outcoupling mirror
as in the coherent mode-locking technique that was proposed in
Ref.~\cite{kozlov}. In conclusion, we may anticipate that few-cycle dissipative
solitons may be also generated in a laser cavity configuration:
such a scheme is reserved for future investigation and will be published
elsewhere.

\acknowledgments

N.N.R. acknowledges the Cariplo Foundation grant of Landau Network -- Centro
Volta for the support of his work at the Universit\'{a} degli Studi di Brescia, 
as well
as the support of the Russian Federal agency on science and innovations,
contract No. 02.740.11.0390, and grants
of the Russian Foundation for Basic Research 09-02-12129-ofi\underline{\hskip2.5mm}m
and of the Russian Ministry of Education
and Science RNP 2.1.1/4694.


\appendix
\section{Derivation of equations (\ref{24}) and (\ref{25})}
In  the first order of the perturbation theory, the wave equation (\ref{8})
takes the form
\begin{equation}
\frac{\partial^2 E_1}{\partial\xi\partial\eta}=-\frac{\pi\sigma_0}{c}\,\frac{\partial
A}{\partial\xi}
\label{A1}
\end{equation}
where $\eta =z+ct$ and $\xi =z-ct$. The general solution of this equation reads
\begin{equation}
E_1(\xi ,\, \eta )=-\frac{\pi\sigma_0}{c}\, E_f(\xi )\eta +F(\xi )+B(\eta )\, ,
\label{A2}
\end{equation}
where $F$ and $B$ are arbitrary functions of their variables. In terms of the
original variables we have
\begin{equation}
E_1(z,\, t)=-\frac{\pi\sigma_0}{c}\, E_f(z-ct)(z+ct)+F(z-ct)+B(z+ct)\, .
\label{A3}
\end{equation}
The first terms in Eqs.~(\ref{A2}) and (\ref{A3}) in the right-hand sides
represent a particular solution of inhomogeneous equation
(\ref{A1}), the second and third terms represent forward ($F$) and backward
($B$) travelling waves, respectively. Finally, the field
$E(z,\, t)$ acquires the form
\begin{equation}
E(z,t)=\left\{
\begin{array}{l}
E_f(z-\tau )+E_b(z+\tau ) \quad\mbox{for}\, z<0\, ,\\
\left[ 1-\frac{\pi\sigma_0}{c}(z+\tau )\right] E_f(z-\tau )+F(z-\tau
)+B(z+\tau ) \quad\mbox{for}\, 0<z<L\, ,\\
E_{tr}(z-\tau )  \quad\mbox{for}\, z>L\, .
\end{array}
\right.
\label{A4}
\end{equation}
Here $\tau =ct$ and $E_f$ is given function -- the shape of the incident pulse.
Unknown functions are $E_f$, $F$, $B$, and
$E_{tr}$; each of them is a function of a single variable. The unknown functions
are to be found from the continuity conditions of $E$
and $\partial E/\partial z$ at $z=0$ and $z=L$:
\begin{eqnarray}
&& E_b(\tau )=-\frac{\pi\sigma_0}{c}\,\tau E_f(-\tau )+F(-\tau )+B(\tau )\, ,
\nonumber\\
&& E_b^\prime (\tau ) =-\frac{\pi\sigma_0}{c}\left[ E_f(-\tau )+\tau
E^\prime_f(-\tau )\right] +F^\prime (-\tau )+B^\prime (\tau )\, ,
\label{A5}\\
&& \left[ 1-\frac{\pi\sigma_0}{c}(L+\tau )\right] E_f (L-\tau )+F(L-\tau
)+B(L+\tau )=E_{tr}(L-\tau )\, ,
\nonumber\\
&& \left[ 1-\frac{\pi\sigma_0}{c}(L+\tau )\right]E_f^\prime (L-\tau
)-\frac{\pi\sigma_0}{c}\, E_f(L-\tau )+F^\prime (L-\tau )+B^\prime
(L+\tau )=E^\prime_{tr}(L-\tau )\, .
\nonumber
\end{eqnarray}
The prime indicates the differentiation of the corresponding function with
respect to $\tau$. Let us rewrite the set of equations
(\ref{A5}) after differentiating the first and the third equations with respect
to $\tau$:
\begin{eqnarray}
&& E^\prime_b(\tau )=-\frac{\pi\sigma_0}{c}\left[ E_f(-\tau )-\tau
E^\prime_f(-\tau )\right]-F^\prime (-\tau )+B^\prime (\tau )\, ,
\nonumber\\
&& E_b^\prime (\tau ) =-\frac{\pi\sigma_0}{c}\left[ E_f(-\tau )+\tau
E^\prime_f(-\tau )\right] +F^\prime (-\tau )+B^\prime (\tau )\, ,
\label{A6}\\
&& -\frac{\pi\sigma_0}{c}\, E_f(L-\tau )-
\left[ 1-\frac{\pi\sigma_0}{c}(L+\tau )\right] E^\prime_f (L-\tau )-F^\prime
(L-\tau )+B^\prime (L+\tau )=-E^\prime_{tr}(L-\tau )\, ,
\nonumber\\
&& -\frac{\pi\sigma_0}{c}\, E_f(L-\tau )+\left[ 1-\frac{\pi\sigma_0}{c}(L+\tau
)\right]E_f^\prime (L-\tau )+F^\prime (L-\tau )+B^\prime
(L+\tau )=E^\prime_{tr}(L-\tau )\, .
\nonumber
\end{eqnarray}
Subtracting the second equation from the first we get
\begin{equation}
F^\prime (-\tau )=\frac{\pi\sigma_0}{c}\,\tau\, E_f^\prime (-\tau )\, ,
\label{A7}
\end{equation}
and therefore
\begin{equation}
F(\tau )=-\frac{\pi\sigma_0}{c}\,\int_{-\infty}^\tau d\theta\, \theta\,
E_f^\prime (\theta )\, .
\label{A8}
\end{equation}
Summing up these two equations we find
\begin{equation}
E^\prime_b(\tau )=-\frac{\pi\sigma_0}{c}\, E_f(-\tau )+B^\prime (\tau )\, .
\label{A9}
\end{equation}

Similar manipulations are to be performed with last two equations in (\ref{A6}).
Summation yields
\begin{equation}
B^\prime (L+\tau )=\frac{\pi\sigma_0}{c}\, E_f(L-\tau )\, .
\label{A10}
\end{equation}
From this result we find
\begin{equation}
B^\prime (\tau )=\frac{\pi\sigma_0}{c}\, E_f(2L-\tau )
\label{A11}
\end{equation}
and
\begin{equation}
B(\tau )=\frac{\pi\sigma_0}{c}\, \int_{-\infty}^\tau d\theta\, E_f (2L-\theta
)\, .
\label{A12}
\end{equation}
Combining equations (\ref{A9}) and (\ref{A11}), we obtain the shape of the
reflected pulse, see Eq.~(\ref{25}).

Finally, by subtracting last two equations in the system (\ref{A6}) from each
other we get
\begin{equation}
E^\prime_{tr}(L-\tau )=\left[ 1-\frac{\pi\sigma_0}{c}(L+\tau )\right]\,
E^\prime_f(L-\tau )+F^\prime (L-\tau )
\label{A13}
\end{equation}
and therefore
\begin{equation}
E^\prime_{tr}(\tau )=\left[ 1-\frac{\pi\sigma_0}{c}(2L-\tau )\right]\,
E^\prime_f(\tau )+F^\prime (\tau )
=\left[ 1-\frac{\pi\sigma_0}{c}\, L\right]\, E_f^\prime (\tau )\, .
\label{A14}
\end{equation}
Integration of the last formula gives Eq.~(\ref{24}).

\end{document}